\documentclass[%
 reprint,
 amsmath,amssymb,
 aps,
]{revtex4-1}

\usepackage{graphicx}
\usepackage{dcolumn}
\usepackage{xcolor}
\usepackage{bm}

\newcommand{\bo}{\raise-1mm\hbox{\Large$\Box$}}
\newcommand{\expva}[1]{\langle #1 \rangle}

\begin{document}


\title{On the spectrum of open strings in the Hard-Wall model of AdS/QCD: the role of $S^5$}

\author{Saulo Diles}
\email{smdiles@ufpa.br}
\affiliation{Campus Salin\'opolis,\\ Universidade Federal do Par\'a,\\
68721-000, Salin\'opolis, Par\'a, Brazil}
\affiliation{Unidade Acadêmica de Física,\\ Univ. Federal de Campina Grande, R. Aprígio Veloso, 58429-900 - Campina Grande}

\date{\today}

\begin{abstract}
We revisit the problem of a classical open string representing a meson in the hard-wall geometry and consider the $S^5$ compact space. The existence of a confining region leads to a discrete set of string configurations parameterized by the number $n$ of complete turnarounds $S^5$ great circles. Unlike the conformal case, the string in the bulk can explore $S^5$ even in the case of equal boundary conditions on its endpoints. The string solutions are parameterized by an internal dimensionless conserved charge $l_n$. The $q\bar{q}$ potential is linear for large separations, and the string tension is sensitive to $l_n$.

\end{abstract}

\maketitle


\section{Introduction}

The AdS/CFT correspondence \cite{Maldacena:1997re, Witten:1998qj} states that the strongly coupled dynamics of a conformal super-symmetric gauge theory in $3+1$ flat space-time is encoded in the ten-dimensional geometry of the $AdS_5\times S^5$ bulk space-time. It is a realization of the holographic principle \cite{Susskind:1994vu, Bousso:2002ju}. Many efforts have been made to use the holographic dictionary of the AdS/CFT correspondence to reveal features of the strongly coupled phenomenology of confining gauge theories like QCD. In this context, confinement is reached as a consequence of the scaling symmetry breaking in the bulk gravity side, introducing a mass scale and a discrete particle spectrum in the dual gauge theory. Confinement is one of the more challenging phenomenologies to be described by theoretical physics. On the one hand, by definition, confined quarks cannot be measured directly in any particle detector, while on the other hand, the interaction of confined quarks is strongly coupled \cite{nambu1976confinement, bander1981theories, braun2010quark}. 
The holographic models obtained by breaking the conformal symmetry in the AdS/CFT correspondence are now known as AdS/QCD, or holographic QCD. The ad-hoc breaking of the conformal symmetry by a hard cut-off due to a D-brane where the bulk spacetime ends was first proposed to study deep inelastic scattering  \cite{Polchinski:2002jw} and then applied to modeling light hadron spectroscopy \cite{Boschi-Filho:2002wdj, Erlich:2005qh, Grigoryan:2007wn}. Such a symmetry breaking can be summarized by cutting the radial direction ($z$) of the bulk space-time, producing the effect of an infinitely rigid wall there.

One important page of the holographic dictionary is established in   \cite{Rey:1998ik, Maldacena:1998im} and maps a pair of interacting particles of the dual gauge theory with a semi-classical open string in the bulk whose endpoints are attached at the conformal boundary. The string endpoints are identified with a quark and an anti-quark according to string orientation, and the hole system is interpreted as a meson state \cite{Erdmenger:2007cm}. In this original version of the AdS/CFT correspondence conformal symmetry constrains the $q\bar{q}$ interaction potential to be of the Coulomb type and there is no confinement in the boundary gauge theory.
In this context, the role of $S^5$ has been discussed either when the string walks on $S^5$ as it goes down in the holographic direction \cite{Maldacena:1998im} and when the role string rotates in $S^5$ as time goes on \cite{Tseytlin:2002tr}. Here we discuss a non-conformal version of the former case, where the rotations in $S^5$ are performed along the string extent, not on time.

We study strings in the hard-wall model of AdS/QCD exploring the $S^5$ sphere in its ten-dimensional bulk space-time geometry of $AdS_5\times S^5$, where the cutoff is placed in the radial direction of the $AdS$ sector. In Ref. \cite{Boschi-Filho:2005nmp} is discussed the Randal-Sundrum construction and probed by open strings, the sphere $S^5$ is absent in this case. Most AdS/QCD models of quark/ anti-quark potentials do not consider the $S^5$ sector of the bulk geometry. Here we focus on the dynamics in $S^5$ and find nice features of the possible string solutions. In particular, we show that equal boundary conditions on $S^5$ at the different endpoints do not lead to a constant angle along the string profile, as it does in the conformal case. The $S^5$ plays a non-trivial role in the confining region: for a given boundary condition on $S^5$ there is a discrete set of admissible separations for the string endpoints, while for a fixed separation in the string endpoints, there are discrete energies in the flux tube corresponding to a discrete set of string tensions.

The article is organized as follows. In Section II, we study the Nambu-Goto action for a string in the complete $AdS_5\times S^5$ geometry of the hard-wall model and solve the equations of motion using conservation laws. In Section III, we discuss the discrete sets of solutions for confined strings and the implications on the physical process of holographically representing a meson in the conformal and confined region. We conclude and discuss future perspectives in Section IV.

 \section{Wilson Lines in Hard-Wall AdS/QCD}{\label{HardWall}}

The holographic prescription maps the spectator value of the Wilson loop operator for $q\bar{q}$ in the boundary into an open string on the $AdS_5\times S^5$ whose endpoints are attached to the boundary. The radial separation $L$ of the $q\bar{q}$ pair is placed by the boundary conditions of the string. The string is static, and the contour for the Wilson loop is a rectangle with infinity length in the time. The Wilson loop operator dual to a string in $AdS_5\times S^5$ is 
\begin{equation}
    \hat{W} =Tr[e^{-\oint \{A_\mu dx^\mu + |dx| \Phi^I\theta^I]\}},
 \end{equation}
where $\theta^I$ are unit vectors on $S^5$ and $\Phi^I$ are scalar fields of the boundary gauge theory.

The expectation value of the Wilson operator relates with the energy in the flux tube  $\expva{\hat{W}}\sim e^{-T U}$. The holographic dictionary states that the expectation value of the Wilson operator of a closed contour in the boundary theory equals the exponential of the on-shell Nambu-Goto action of the string $\expva{W}= e^{-S_{_{NG}}},$ where for a string parameterization $X^m(\sigma^\alpha)$:
\begin{equation}
    S_{NG} = \frac{1}{2\pi\alpha'}\int d^2\sigma \sqrt{det(g_{mn}\partial_\alpha X^m\partial_\beta X^n)}.
\end{equation}
For a static configuration, time factorizes, and the interaction energy is given by the pure space integral.

We adopt Poincaré patch  to describe the Euclidean $AdS_5,$ with Minkowski coordinates $(t,\vec{x})\in \mathcal{R}^{1,3},~z\in (0,z_{hw}]$ and  the standard angle element for the five-dimensional sphere \cite{Boschi-Filho:2001qcm}:
\begin{equation}
 ds^2 = \frac{R^2}{z^2}(dt^2 + d\vec{x}^2 + dz^2) + R^2d\Omega_5^2.
\end{equation}
The sector of the geometry corresponding to the $S^5$ sphere posses a positive definite metric implying that any displacement there increases the string Lagrangean and, as a consequence, the string walk on the geodesics of $S^5$ connecting the locations where string endpoints are attached.  The geodesics along $S^5$ are great circles parametrized by the angles $\theta \sim \theta + 2\pi.$ We  parameterize the world-sheet as $X^m(t,x)= (t,x,0,0,z(x),\theta(x),0,0,0,0)$, and set the boundary conditions as $z(x\to\pm L/2)\to 0$ and $\theta(x\to\pm L/2)\to \pm \delta\theta_0/2$. 

The $S^5$ boundary conditions obeyed by the dual semi-classical string on the bulk space-time are associated with the breaking of the $U(N)$ symmetry in the boundary due to a Higgs mechanism giving values for one scalar field $\vec{\Phi}$ for each symmetry breaking, providing a Dirichlet boundary condition for  $\vec{\theta}\in S^5,~\vec{\theta}\equiv\frac{\vec{\Phi}}{|\vec{\Phi}|}$, at the string endpoint. Non-trivial boundary Dirichlet conditions arise in the case  $U(N)\to U(N-2)\times U(1) \times U(1)$. We remark that the original proposal of AdS/CFT correspondence requires the inclusion of the $S^5$ degrees of freedom for the stretched string representing a quark/anti-quark pair \cite{Aharony:1999ti}. A detailed discussion of the boundary conditions of the bulk string on $S^5$ is found in \cite{Drukker:2005cu}, observing that in this reference it is used the Polyakov action. The case of non-trivial boundary condition on $S^5$ is discussed in the original proposal where conformal symmetry is preserved \cite{Maldacena:1998im}.

In the present case, the  Nambu-Goto action takes the same form as the conformal one, given by
 \begin{equation}
 S_{NG}=T\frac{R^2}{2\pi\alpha'} \int_{-L/2}^{L/2} dx \sqrt{\frac{1+z'^2+ z^2\theta'^2}{z^4}}. \label{actions5}
\end{equation}
For simplicity we set  $\frac{R^2}{2\pi\alpha'}=1$ and the 
 action is expressed as $S_{NG} = T\int dx \mathcal{L}(z,z',\theta,\theta';x)$, with a Lagrangean density 
 \begin{equation}
 \mathcal{L}=\sqrt{\frac{1+z'^2+ z^2\theta'^2}{z^4}}.
 \end{equation} 
 We note that the Langrangean density is $\theta$ independent, leading to the conservation of the momentum $p_\theta = \frac{\partial\mathcal{L}}{\partial \theta'}$ and a first order equation for $\theta$:
 \begin{equation}
  \frac{d\theta}{dx} = \pm p_\theta\sqrt{\frac{1+z'2}{1-p_\theta^2 z^2}}. \label{ang}
 \end{equation}
Note that $p_\theta$ is bounded from above. Reality of $\theta'$ imposes a imposes the restriction $p_\theta<1/z_{\textrm{max}}$, where $z_{\textrm{max}}$ is the turning point of the world-sheet. In the hard-wall geometry, we have a global maximum defined by $z_{\textrm{hw}}$ imposing for any string that  $p_\theta<\frac{1}{z_{\textrm{hw}}}.$
 In eq.\eqref{ang}, the single choice corresponds to the orientation of the rotation, and we will always choose orientation so that the total angle varied by the string satisfies $\delta\theta_0\leq\pi$.

The Lagrangian density does not depend explicitly on $x$, and there is a conserved Hamiltonian 
   $\mathcal{H}= \mathcal{L} -  z'\frac{\partial \mathcal{L}}{\partial z'} - \theta'\frac{\partial \mathcal{L}}{\partial \theta'}$ that is given by
 \begin{equation}
\mathcal{H}=  \frac{1}{z^2\sqrt{1+z'^2+z^2\theta'^2}}.  \label{hamilton}
 \end{equation}
We obtain a simple expression for $z'$:
\begin{equation}
 \frac{dz}{dx} = \pm \frac{\sqrt{1-z^2 p_\theta^2 - \mathcal{H}^2z^4}}{\mathcal{H}z^2}, \label{dist}
\end{equation}
where the plus sign applies to the region where the string moves away from the boundary and the minus sign where the string approaches the boundary. The cut-off in the geometry is set along the radial direction and does not break the reflection symmetry around $x=0$. 
There is a critical separation for the boundary $q\bar{q}$, $L_c$, such that the string bottom reaches the wall. Continuity of the first derivative $z'(x)$ requires the wall location to coincide with the point of return: $z'|_{z_{\textrm{hw}}}=0$. It leads to
 \begin{equation}
  \mathcal{H} = \frac{\sqrt{1-z^2_{\textrm{hw}}p_\theta^2}}{z_{\textrm{hw}}^2},
 \end{equation}
and we find  for a counterclockwise rotation: \begin{equation}
  \frac{d\theta}{dx} = \frac{p_\theta}{\mathcal{H}z^2}. \label{newang}
 \end{equation}

The critical separation $L_c$ where the bottom of the string reaches the wall is obtained fixing $z_{\textrm{max}}=z_{\textrm{hw}}$:
\begin{equation}
\begin{aligned}
    L_c(l) &= 2z_{\textrm{hw}}\sqrt{1-l^2}I_1(l), \\
    I_1(l)&= \int_0^1dy \frac{y^2}{\sqrt{1-l^2y^2-(1-l^2)y^4}},
\end{aligned}  
\end{equation}
 where we define the dimensionless conserved charge $l\equiv z_{\textrm{hw}}p_\theta$ with range $l\in(0,1]$ associated with the worldsheet path along $S^5$. The existence of a critical length separates the Coulomb phase from the linear confinement, and results in a dual $q\bar{q}$ potential of the type
 \begin{equation}\label{potential}
     V(L)=\begin{cases}
 -\frac{a}{L}, & 0< L <L_c(l), \\
\sigma L, & L\geq L_c(l).
\end{cases}
 \end{equation}
This is the typical regularized potential of the hard-wall model. This type of potential was previously used to model the asymptotic of the heavy quarkonium interaction \cite{Ono:1979tk}.
 
Assuming $z_{\textrm{hw}}$ posses dimensions of length, $l$ is a dimensionless parameter while $p_\theta$ possesses a dimension of $\frac{energy}{\textrm{lenght}}$ which we associate with an angular momentum since we treat the $x$ coordinate formally as a time in the Lagrangian formulation of the world-sheet action. We refer to the parameter $l$ as the angular momentum of the string everywhere in this work. In the above equation and the forthcoming integrals, the lower bound corresponding to $z\to 0$ must be interpreted as $z= \epsilon$. After removing any $\frac{1}{\epsilon}$ divergences we perform the limit $\epsilon \to 0$. 
 
For a shorter separation, the wall is irrelevant, but for separations larger than $L_c$ the wall plays the fundamental role. Since $\theta'$ decreases on $z$, the wall location is a global minimum of the action. The exceeding part of the string lies on the wall where it holds $z'=0$.  If there is no wall larger $L$ corresponds to a large value $z_{\textrm{max}}$ for the location of the bottom of the string while in the hard-wall model the space ends at $z_{\textrm{hw}}$, and therefore $z_{\textrm{max}}\leq z_{\textrm{hw}}$. Continuity in $z'(x)$  is natural from classical mechanics, in this case, it implies a piece-wise structure of the worldsheet. For $L>L_c$ the worldsheet is composed of two external pieces fitting a "pure AdS solution" connected at the points $x_\pm = \pm \frac{L-L_c}{2}$ with a central part which lies at the wall. Continuity of $z'(x)$ at the connections requires the "pure AdS solution" to be the one with $z_{\textrm{max}}=z_{\textrm{hw}}$. Such configuration is continuous, and its first derivative is continuous but non-analytical. Second derivative of $z(x)$ is discontinuous at  $x_\pm = \pm \frac{L-L_c}{2}$. For example, in the limit $x\to -\frac{L-L_c}{2}$ we find from the left $z''<0$ while from the right $z''=0$. 

If we restrict to solutions with no angular momentum, or $l=0$, we recover the profile obtained in Ref.\cite{Boschi-Filho:2005nmp}. In this reference, the profile is obtained from minimal energy arguments. 

Let's focus on the walk along the great circle of $S^5$ in the confining region ($L>L_c$). We have from eq.(\ref{newang}) that at $z_{\textrm{hw}}$ the angular velocity assume a constant value:
\begin{equation}
    \theta'|_{z_{\textrm{hw}}}= \frac{p_\theta}{\mathcal{H}z_{\textrm{hw}}}.
\end{equation}
The total angle varied along the hole string is given by
 \begin{equation}
      \delta \theta =2\int_0^{z_{\textrm{hw}}} dz \frac{p_\theta}{\sqrt{1-z^2 p_\theta^2 - \mathcal{H}^2z^4}} +  \frac{p_\theta}{\mathcal{H}z_{\textrm{hw}}^2}(L-L_c).
 \end{equation}
It is convenient to write the above equation in the form:
\begin{equation}\label{circle}
 \delta\theta(L,l) = l I_2(l)   + \frac{l}{z_{\textrm{hw}}\sqrt{1-l^2}}(L-L_c(l)),
\end{equation}
where 
\begin{equation}
 I_2(l) = \int_0^1\frac{dy}{\sqrt{1-l^2y^2-(1-l^2)y^4}}.
\end{equation}

\section{The physics of the discrete string spectrum}

 For $q\bar{q}$ separations above the critical, the internal symmetries of the worldsheet lead to degenerate classical configurations that follow from Dirichlet boundary conditions on $S^5$. We remember that the boundary conditions on $z\to0$ are imposed by taking a slice infinitesimally close to the boundary at some $z=\epsilon$,  setting boundary conditions there, and, at the end of the day, after regularizing the integrals, we take the limit $\epsilon\to 0$. 

 The geometry of the hard-wall model in AdS/QCD classifies the strings in two different regions: the conformal and the confining. The dimensionless conserved charge associated with internal rotations, $l$,  fixes the critical $q\bar{q}$ separation at the boundary. The role imposed by the discrete set of solutions expressed in eq.\eqref{discrete} can be discussed in two different cases. The first one corresponds to the meson that is created in the conformal region, and the second situation corresponds to the creation of a meson in the confined region.  

In the first situation, the holographic string can be created with \textit{any} value of angular momentum $l$, say $l_0\in [0,1)$. The angular momentum $l_0$ define the angular ($S^5$) boundary condition and the critical separation $L_{c,0}\equiv L_c(l_0)=2z_{hw}\sqrt{1-l^2}I_1(l_0)$. The physical process of separating the pair corresponds to changing the boundary conditions on the $x$ coordinate and it should not change the boundary conditions on $S^5$. In the mathematical sense, we are looking for the complete set of solutions with $\delta \theta=\delta \theta(l_0)$ fixed. While on the physical side, we remember that $l$ is a conserved charge born from a continuous symmetry, and we expect it to keep constant in any physical process, which is the case of changing string endpoint separation. 

While in the conformal region the $q\bar{q}$ pair can be continuously separated up to the critical distance $L_c$. When the pair reaches the critical separation it enters the confined region and this is no more the case. In the confining region, the separation cannot increase continuously without changing angular boundary conditions. So, it can change only discretely. The allowed step is fixed by $l$ to be one solution of
\begin{equation}\label{Ln}
L_n(l) = L_c(l)+n\times\left(\frac{2\pi z_{hw}\sqrt{1-l^2}}{l}\right).
\end{equation}
The constituent quarks at the string endpoints are constrained in a lattice with lattice spacing \begin{equation}\label{latice}
    a(l)=\frac{2\pi z_{hw}\sqrt{1-l^2}}{l},
\end{equation}
for a non-vanishing $l$. In this sense, the presence of the $S^5$ in the holographic dictionary tells us that confinement puts the constituent quarks of meson into a lattice. Non-trivial boundary conditions on $S^5$ lead to a $l\neq0$ and fix a non-vanishing lattice space. If we fix $l=0$  eq.\eqref{circle} imposes $\Delta\theta=0$ independent of $L$ and the set of locations one can put the string endpoints is continuous. The lattice spacing decreases monotonically with angular momentum, and the continuum space is recovered in the limit $l_0\to 1$.

The interaction energy of the $q\bar{q}$ is proportional to the regularized Nambu-Goto action. As usual, regularization means subtracting the action of two strings attached to a single point in the boundary and stretching down the AdS space. In our case, each single string ends at the wall and represents the infinity mass of the constituent quark.

The energy of the $q\bar{q}$ pair is given by the action integral, which is divergent in the boundary. The regularization is the usual, subtracting the action of two strings stretching from the quark location in the radial AdS direction. We have in the conformal region
\begin{align}
E(L,l) &= \frac{\sqrt{1-l^2}I_2(l)}{L}\int_0^1dy\frac{1}{y^2}\times \nonumber \\
&\left(\frac{1}{\sqrt{1-l^2y^2-(1-l^2)y^4}} -\frac{1}{\sqrt{1-l^2y^2}} \right),    
\end{align}
with $L\in(0,L_c(l))$. In the confining region the configuration space is discrete, $L(m)=L_c(l)+ma(z_{\textrm{hw}},l),~m=1,2,3,...$, and we have discrete energy levels
\begin{equation}\label{confEnergy}
\begin{aligned}
 E(m,l)  &= E_c + \sigma (L(m)-L_c) \\
 &=  E_c + m\left(\frac{a(z_{\textrm{hw}},l)}{z_{\textrm{hw}}²\sqrt{1-l^2}}\right),
\end{aligned}
\end{equation}
where the critical energy is $E_c=E(L_c(l),l)$ and we define 
 \begin{equation}
     \sigma \equiv \mathcal{L}_{NG}|_{z_{\textrm{hw}}}
= \frac{1}{z_{\textrm{hw}}^2\sqrt{1-l^2}}.
\end{equation}

\begin{figure}
  \includegraphics[width=2.5 in]{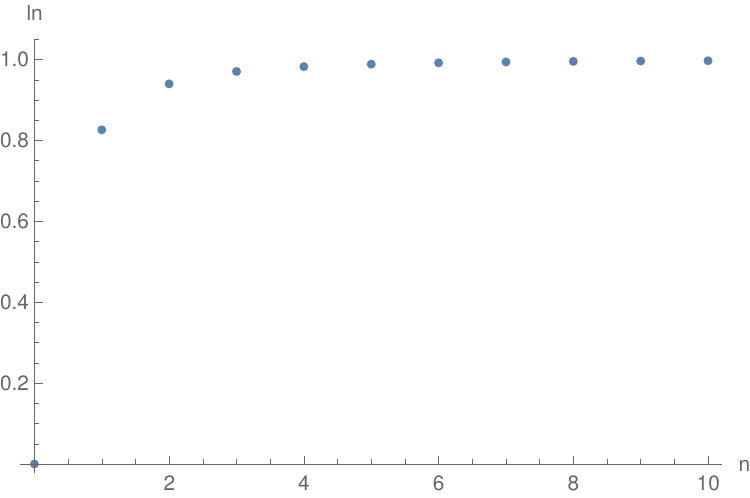}
\caption{Discrete values for the internal angular momentum of a string created with endpoint separation $L>L_c$ in the boundary. This spectrum correspond to $z_{\textrm{hw}}=1$ and $L=5$. }
\label{fig:quantum}
\end{figure}

The second situation is when the quark/anti-quark pair is created in the confining region. In this case, the analysis is subtle. For a given boundary condition $\delta\theta\in(0,\pi)$ on the great circle of $S^5$, there is an angular momentum $l_0$ which in the conformal region defines a critical length $L_0\equiv L_c(l_0)$. Provided that the pair is created with separation $L>L_0$ we note two facts: i) The critical length is monotonically decreasing with $l$, which imply that if the pair is created with an angular momentum $\tilde{l}>l_0$ this string still belong to the confining region, and ii) The boundary condition on $\delta \theta_0$ allows for a finite number of complete turn around $\delta\theta = \delta \theta_0 +n(2\pi)\sim \delta\theta_0$. However, a string with momentum $l_0$ walks $\delta\theta_0$ in the piece of the string apart from the wall. Therefore to reach the same $\theta_0$ the string has been walked around the $S^5$ great circle at least one time, imposing $n\geq 1$. These two facts put together mean that when a string is created in the confining region, its angular momentum belongs to the set of solutions of the algebraic equation:
\begin{equation}\label{discrete}
 \delta\theta_0 + n(2\pi) =     l_n I_1(l_n)   + \frac{l_n}{z_{hw}\sqrt{1-l_n^2}}(L-2z_{hw}I(l_n)).
\end{equation}
 In this case, the minimal momentum $l_0$ corresponding to no complete turnaround is forbidden.   One set of the discrete spectrum of $l_n$ is represented in Fig.\ref{fig:quantum}.

It is important to remark that  $\delta\theta=0$, which corresponds to the trivial boundary conditions on $S^5$, also leads to the discrete spectrum of angular momentum. The existence of world-sheet solutions with $\delta\theta_0=0$ and $l=l_n\neq 0$, the roots of eq.\eqref{discrete}, shows clearly that the problem in the full geometry $AdS_5\times S^5$ of the AdS/CFT correspondence is not equivalent to the problem restricted to the $AdS_5$ sector when we place the hard-wall. In this case, where conformal symmetry is broken, the problem with trivial boundary conditions on $S^5$ is not equivalent to suppressing the $S^5$ sphere in the bulk space geometry.

For each value of $n$, there is a different critical length and a different contribution for the linear part of the potential. Moreover, once one of the $\{l_n,n=1,2,3...\}$ is chosen, the critical length is fixed, and the lattice of space distance is given by eq.\eqref{latice}. 

\begin{figure}
  \includegraphics[width=2.5 in]{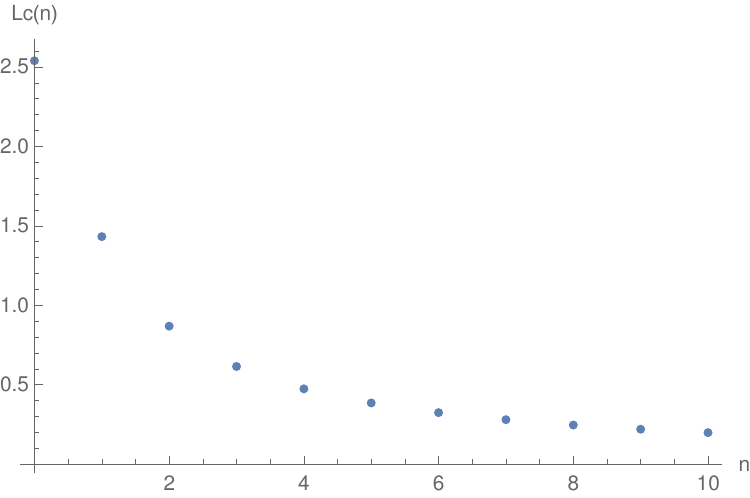}
\caption{Critical $q\bar{q}$ separation as a function of the widding number $n$ for $z_\textrm{hw}=1$. }
\label{fig:critical}
\end{figure}

 The critical $q\bar{q}$ separation defining the confining region is also $l$ dependent $ L_c(l) = 2z_{\textrm{hw}}I_1(l)$
so there is a discrete set of critical separations, corresponding to the discrete set of allowed angular boundary conditions. We represent the $n$ dependence of $L_c$ in Figure \ref{fig:critical}.

\subsection{$n$ is a winding number}

 The existence of a cut-off in the radial AdS part of the $AdS_5\times S^5$ geometry generates linear confinement in the sense of the quark anti-quark interaction potential. For classical strings in the confining region, the angular momentum associated with the $S^5$ sphere assumes discrete values. For a given separation $L>L_c$, the angular velocity of the string along the great circle of $S^5$ is constant. The discrete values of the angular momentum are a consequence of fixing Dirichlet boundary conditions in angular coordinates, and the discrete index counts precisely the number of string turnarounds the $S^5$ great circle in the confining region. So, the $n$ in eqs.\eqref{Ln} and \eqref{discrete} are precisely a winding number or a topological charge. The classical mechanics of this problem leads to a local geometry in the confining region of the cylindrical type: the string walks around a $S^1$ circle by the time it walks along a straight line $\mathbb{R}$ at the wall. The local geometry felled by the string is $\mathbb{R}\times S^1$, which produces the topological charge discretizing the spectrum of classical strings. 

\subsection{The position space for the dual infinitely heavy quark and anti-quark pair}

The hard cut-off imposes a critical value $L_c$ for the quark/ anti-quark separation: for $L>L_c$ we are in the confining region. Worldsheets representing confined mesons are necessarily non-analytic, there is an unavoidable discontinuity in the second derivative of its parameterization. However, we have a well-defined classical mechanic description that requires continuity of the functions that parameterize the worldsheet and of its first derivative. It is a necessary and sufficient condition to obtain a suitable configuration respecting the requirements of a Lagrangian theory.  We observe that the non-trivial boundary condition on $S^5$ requires a non-vanishing ($l\neq 0$), which implies that the string solutions in the confining region possess its endpoints restricted to the discrete set of solutions where the angular boundary conditions are compatible with the longitudinal boundary conditions. For a given $l$ there is a discrete set of compatible values $L_n>L_c$ to put the constituent quarks. This position space is precisely a lattice with lattice spacing defined by eq. \eqref{latice}. In this sense, a non-vanishing $l$ sets the quarks into a lattice space. 

In the particular case where string endpoints reach the same point of $S^5$ in the boundary allow for $l=0$ solutions, and in this case, there is no restriction in the space boundary condition and the $q\bar{q}$ pair can have any separation. However, this is not the only possibility. There is a discrete set of worldsheets with non-vanishing $l$ defined by the solutions of eq.\eqref{discrete} with $\delta\theta_0=0$. In this set of worldsheets, the space boundary conditions also get discretized.

A strong criticism is faced when we intend to interpret this mathematical fact as a connection between bottom-up holography and lattice QCD  \cite{kogut1983lattice, Ratti:2018ksb}. On one hand, it is a natural association as far as we interpret our string endpoints as representing quark locations. On the other hand, the holographic model describes infinitely heavy quarks that require an infinity amount of energy to be moved. In this sense, the process of continuously separating the quark away from its anti-quark is nonphysical in the holographic hard-wall model. So, we conclude that a sensible physical interpretation of our results is that for strings representing confined $q\bar{q}$ pairs: i) boundary conditions on $S^5$ and $\mathbb{R}^{3}$ are not independent and for each angular boundary condition there exists a lattice set of space boundary conditions; and ii) for a given $q\bar{q}$ separation and a given boundary condition on $S^5$ there exists a discrete set of classical strings in the bulk parameterized by $l_n$. The $q\bar{q}$ pair can be placed with any space separation but their dual description in terms of a string in the bulk geometry of the hard-wall model is not unique.

\subsection{The string tension is $n$ dependent}

 The confining signature of the  $q\bar{q}$n pair appears in its interaction potential as the linear term in the Cornell potential \cite{eichten1978charmonium}. In the holographic hard-wall model this therm is a consequence of the hard cut-off in the bulk geometry: when the string reaches this region it just sits on the wall, and the string action is linear and given by the local Lagragean times the length of the string along the wall. In the case of the hard-wall model, the Coulomb term and the linear term are mutually excluded. 

The definition of the string tension is the limit $\lim_{L\to \infty}\frac{E}{L}$.  The relation between energy and distance is linear so that the $\sigma$ in eq.\eqref{confEnergy} is precisely the string tension. The existence of a discrete set of strings in the bulk compatible with the same $q\bar{q}$ configuration in the boundary implies that $n$ also parameterizes the possible values of the energy density inside the flux tube. This energy is proportional to the total string action and the length of the string along the wall. A larger $n$ means that there are more turnarounds on $S^5$ as the string walks in the hard-wall, and the effective length of the string there gets larger. The more turnarounds per length in the $x$ direction leads to a large contribution to the action integral, which is holographically mapped into a large amount of energy per $q\bar{q}$ separation.
It translates into the analytic result for the string tension: 
\begin{equation}
\sigma_n = \frac{1}{z_{\textrm{hw}}^2\sqrt{1-l_n^2}},    
\end{equation}
 with the possible $l_n$ are the solutions of eq.\eqref{discrete} which are increasing in $n$ with asymptotic value $l_n\to1^-$ as $n\to \infty$. The possible string tensions are bounded from below with a minimum defined by $z_{\textrm{hw}}$, $\sigma_0= \frac{1}{z_\textrm{hw}^2}$, but unbounded from above.

\section{Conclusions}
The present work discusses the role $S^5$ plays in the set of worldsheet solutions representing $q\bar{q}$ pairs in the holographic hard-wall model. 
In the conformal region, nothing changes since the region of the bulk space probed by the strings does not feel the presence of the wall. In the confining region, the $S^5$ sector of the geometry becomes very important and classifies the string solutions into discrete sets where the boundary conditions at the transverse space coordinates of the $AdS_5$ and the $S^5$ are not independent. The solutions are characterized by the total number of turnarounds in $S^5$ along the part of the string sitting at the wall, which define the possible values of $l_n$ as given by the solutions of eq.\eqref{discrete}.  

The compactness of $S^5$ is responsible for the degeneracy of the solutions for a given boundary conditions provided the string can wind it as many times as the boundary conditions allow. The more the string winds the sphere the larger the string action and consequently the larger the energy density inside the flux tube of the dual $q\bar{q}$ pair. It translates in the monotonic growth of the string tension with the winding number $n$, divergent in the $n\to\infty$ limit. In this sense, the winding number works as a principal quantum number in a sense analogous to the Bohr-Sommerfeld model. The topology of $S^5$ allows for a string winding it to shrink. In the present discussion, it could be associated with a decay of the open string in an excited state by emitting a closed string and reaching a fundamental state where the string wind $S^5$ as little as possible. This kind of transition is usually a result of the instability of the solutions under small perturbations. The issue of stability and string decay emerges naturally in our discussion and is a subject for future investigation.

We close this section by speculating about the generalization of the present results to other holographic models for confinement. We expect that open strings in holographic models such as D3/D7 \cite{Karch:2002sh}, Sakai-Sugimoto \cite{Sakai:2004cn} and soft-wall like  \cite{Karch:2006pv, Andreev:2006vy, Andreev:2006ct, Bruni:2018dqm, Diles:2018wbe} also encodes a non-trivial behavior when we allow the worldsheet to explore the compact space. In our present case, the string configuration is exactly determined with a well-defined location for the wall. We expect that the general statement proven in Ref.  \cite{Kinar:1998vq} concerning the criteria for linear confinement in holographic models puts a light on this subject. Such a result has been successfully used to prove a general statement concerning the drag force in the confining vacuum \cite{Diles:2019jkw}. In general, the holographic models of confinement implement an effective wall in the bulk space preventing the strings representing mesons from exploring the infrared region of the bulk space-time. We expect that our present results hold in a good approximation in any confining holographic model by defining an effective wall location $z_{\textrm{hw}}$, while exact results would require a precise mathematical treatment which is out of the scope of the present paper. Moreover, the emergence of the wind number $n$ is a consequence of the compactness of $S^5$ and we expect that a similar situation will be faced in the cases where the bulk geometry includes an arbitrary compact space.

\section{Acknowledments}
Saulo Diles thanks Nelson Braga, for the original motivation of the present work. Saulo Diles also thanks to the Conselho Nacional de Desenvolvimento Científico e Tecnológico (CNPq), Brazil, Grant No. 406875/2023-5.

\bibliographystyle{ieeetr}
\bibliography{apssamp.bib}
  
\end{document}